# Modeling and analysis of driver behavior under shared control through weighted visual and haptic guidance

Zheng Wang[a]*, Rencheng Zheng[a], Edric John Cruz Nacpil[a] and Kimihiko Nakano[a]

[a]*Institute of Industrial Science, The University of Tokyo, Tokyo 113-8654, Japan*

Corresponding author contact information: Zheng Wang, z-wang@iis.u-tokyo.ac.jp

Funding: This work was supported by a Grant-in-Aid for Early-Career Scientists (no. 19K20318) from the Japan Society for the Promotion of Science.

# Modeling and analysis of driver behavior under shared control through weighted visual and haptic guidance

For the optimum design of a driver-automation shared control system, an understanding of driver behavior based on measurements and modeling is crucial early in the development process. This paper presents a driver model through a weighting process of visual guidance from the road ahead and haptic guidance from a steering system for a lane-following task. The proposed weighting process describes the interaction of a driver with the haptic guidance steering and the driver's reliance on it. A driving simulator experiment is conducted to identify the model parameters for driving manually and with haptic guidance. The proposed driver model matched the driver input torque with a satisfactory goodness of fit among fourteen participants after considering the individual differences. The validation results reveal that the simulated trajectory effectively followed the driving course by matching the measured trajectory, thereby indicating that the proposed driver model is capable of predicting driver behavior during haptic guidance. Furthermore, the effect of different degrees of driver reliance on driving performance is evaluated considering various driver states and with system failure via numerical analysis. The model evaluation results reveal the potential of the proposed driver model to be applied in the design and evaluation of a haptic guidance system.

## 1. Introduction

Over the past decade, owing to the continuous progress in the development of vehicle automation, an increasing amount of research has focused on advanced driver assistance systems (ADAS). These systems have been designed to improve both driver safety and comfort. The adaptive cruise control (ACC) systems (Van Arem et al. 2006) and lane keeping assistance (LKA) systems (Pilutti and Ulsoy 1999) among others have been developed for enhancing longitudinal and lateral driving performance.

To further improve driver safety and comfort, driver-automation shared control during a driving task (i.e., combining the abilities of human driver and the vehicle automation) has been investigated (Flemisch et al. 2003, Abbink et al. 2018, Griffiths and Gillespie 2004). Haptic guidance steering control is a type of driver-automation shared control that enables the driver and the automation system to simultaneously control the steering wheel through a haptic interface (Mulder et al. 2008, Forsyth and MacLean 2006). In such a case, the driver is influenced not only by the visual guidance from the road ahead but also by the haptic guidance from the steering wheel. However, the authority of the driver is always higher than that of the automation system because the driver can always overrule conflicting automated steering wheel movements. Thus, a haptic guidance steering system is designed to merely guide the driver along the target trajectory using assistive torques on the steering wheel (Forsyth and MacLean 2006, Wang et al. 2017b).

Ideally, when driving with a haptic guidance system, the driver can comfortably rely on the haptic guidance torque to drive more safely. It has been found that haptic guidance can assist drivers in performing appropriate actions for curve negotiation by producing both the recommended direction and magnitude of a suitable steering operation (Mulder et al. 2012). Haptic guidance systems have also been developed for continuously and smoothly supporting lane-changing maneuvers while maintaining the benefits of the lane-keeping system (Tsoi et al. 2010). An active steering system has also been designed to assist drivers in maneuvering the steering wheel promptly and steadily for emergency obstacle avoidance (Iwano et al. 2014). Haptic guidance steering also leads to a reduced workload as demonstrated by faster reaction times and better secondary task performance (Griffiths and Gillespie 2005). When drivers are fatigued or distracted, the haptic guidance steering reduces their lane departure risk (Wang et al. 2017b, Blaschke et al. 2009). When visual information from the road ahead is degraded as in the case of dense fog, haptic guidance steering delivers a better lane-following performance than that offered by manual driving (Wang et al. 2018, Griffiths and Gillespie 2005, Forsyth and MacLean 2006).

However, haptic guidance systems also have drawbacks. Some negative results of haptic guidance steering have been noted; these include an increased number of collisions because the system does not guide the driver around a road obstacle within a lane (de Winter and Dodou 2011). Such a case demands an evasive maneuver by the driver, and the driver may have to counteract the haptic guidance torque if the system maintains the vehicle along the center of the lane (Griffiths and Gillespie 2005). It is also possible that the haptic guidance system provides inaccurate information in other circumstances. In curve negotiation, for example, the system may misestimate the radius of a curve, thereby resulting in inappropriate steering torque (de Winter and Dodou 2011). In the above cases, the driver has to commit more effort as compared to manual driving or fully-automated driving. Furthermore, the conflict that occurs between the haptic guidance torque and driver's torque can induce a feeling of intrusion in the driver (Lefèvre et al. 2015).

Given these disadvantages, a better understanding of driver behavior based on measurements and modeling is expected to lead to better-designed haptic guidance systems that contribute to the improvement of driver safety and comfort. In order to model driver behavior, many attempts have been made in a mathematical way; these include control theoretic model (Hess and Modjtahedzadeh 1990, Mars and Chevrel 2017), fuzzy control (Xi and Qun 1994), stochastic methods (Qu et al. 2014), and machine learning methods (Martinez-Garcia et al. 2016).

In that regard, the current understanding of driver behavior under driver-automation shared control through a haptic steering interface is limited. Given that the human brain is thought to weigh and integrate multisensory guidance (Ernst and Banks 2002, Serwe et al. 2011), our aim is to propose a driver model that integrates visual guidance from the road ahead and haptic guidance from the steering system. It is hypothesized that a driver relies on visual and haptic guidance through a weighting process based on their respective reliabilities. It is expected that the proposed weighting process is capable of describing driver interactions and reliance in haptic-guidance steering through a small number of parameters in a driver model. This can be beneficial in energy saving when performing online driver model identification for a real-vehicle development process.

Experimental data are acquired from a driving simulator study (Wang et al. 2019) where the subjects were asked to keep the vehicle at the centerline of the lane as accurately as possible, both when driving manually and with haptic guidance. This study investigated the relationships between driver steering and gaze behavior under shared control.

This paper is organized as follows. Section 2 introduces a comprehensive driver-automaton shared control model. Section 3 presents a driving simulator experiment including the participants, apparatus, and scenario. Section 4 illustrates the model identification, validation process, and results. Section 5 evaluates the model by using a numerical simulation to test the ability of the proposed model to predict driver behavior. Discussions are addressed in Section 6, and conclusions and future work are described in Section 7.

## 2. Driver-Automation Shared Control Model

The driver-vehicle-road system is a broad research topic that considers complicated driving situations and numerous driver and environmental factors in a driving task. In this section, the modeling of a driver-vehicle-road system focuses on a lane-following steering task.

### *2.1. General structure*

Fig. 1 shows a general structure of a driver-vehicle-road model with haptic guidance steering system. The driver-vehicle-road model consists of a system that includes a visual system, a neuromuscular system, a haptic guidance system, a steering system, vehicle dynamics, and a road path, which is inspired by (Saleh et al. 2013, Mars et al. 2011). The haptic guidance system provides a haptic guidance torque, $T_h$, to the steering system; a manual driving condition accords with $T_h = 0$. The systems are discussed in detail below.

### *2.2. Vehicle dynamics*

If a vehicle travels without rolling at a constant speed, the vehicle vertical height can be neglected, and only the lateral and yaw motions must be considered. The vehicle is represented as a rigid body projected on the ground. As shown in Fig. 2, in order to simplify vehicle motion analysis, a four-wheeled vehicle can be transformed into an equivalent two-wheeled vehicle or bicycle model given that the left and right tire side-slip angles are equal, the steer angle is small, and there is negligible roll. Table 1 presents the variables and parameters of the equivalent bicycle model.

The linearized vehicle dynamics for describing the vehicle motion in the x-y plane are provided as follows (Masato 2009):

$$mv\frac{d\beta}{dt} + 2(K_f + K_r)\beta + \left\{mv + \frac{2}{v}(l_f K_f - l_r K_r)\right\}r = 2K_f\delta \quad (1)$$

$$2(l_f K_f - l_r K_r)\beta + I\frac{dr}{dt} + \frac{2(l_f^2 K_f + l_r^2 K_r)}{v}r = 2l_f K_f\delta \qquad (2)$$

The vehicle dynamics linked with the steering column dynamics through a self-aligning torque are expressed as

$$T_a = K_{aln} \left( \beta + \frac{l_f r}{v} - \delta \right) \qquad (3)$$

$$K_{aln} = 2E_t K_f K_t \left( \frac{1}{1 + 2E_t K_f / K_s} \right) \qquad (4)$$

### 2.3. Steering column dynamics

An automotive steering column is a device intended primarily for connecting the steering wheel to the steering mechanisms. The variables and parameters of a steering column are presented in Table 1. As the steering wheel is simultaneously actuated by the driver input torque, haptic guidance torque, and aligning torque, the steering wheel system dynamics is expressed as follows:

$$J_s \ddot{\varphi} + B_s \dot{\varphi} = T_d + T_h + T_a \qquad (5)$$

Here, $J_s$ and $B_s$ represent the steering system inertia and damping, respectively. Additionally, a transmission ratio between the steering wheel angle and front wheel angle is described as

$$\delta = K_t \varphi \qquad (6)$$

### 2.4. Visual perception geometry

Fig. 3 shows the geometric relationship between the road and a driver's two-point visual model. The target trajectory is set by the centerline of the lane. This driver's visual model relies on a near and a far point from the road ahead; the near point is used to maintain a central position within the lane, and the far point is used to predicting the curvature of upcoming road. The lateral error, $e_y$, at the near point is defined as the distance from the target trajectory. The yaw error, $e_\theta$, at the far point is defined as the angle between the vehicle movement direction and target trajectory. A proportional-integral controller managing $e_y$ and a proportional controller managing $e_\theta$ for the driver's two-point visual model have been proposed (Salvucci and Gray 2004).
Therefore, the target steering wheel angle, $\varphi'$, can be expressed as

$$\varphi'(t + t_p) = a_1 e_y + a_2 \int e_y dt + a_3 e_\theta \qquad (7)$$

Here, $t_p$ represents the processing time delay by the human driver for a decision making (Masato 2009).

### 2.5. Driver model

The proposed driver model shown in Fig. 1 consists of a visual system and a neuromuscular system. The variables and parameters of the driver model are shown in Table 2. The target steering wheel angle $\varphi'$ is converted into a driver input torque $T_d$ via the neuromuscular system. $K_d$ represents the steering torque provided by the neuromuscular system in proportion to $\varphi'$. $K_{nms}$ represents the neuromuscular reflex gain that resists the disturbance caused by external torque on the steering wheel. Thereby, $K_{nms}$ continually serves to diminish the difference between $\varphi$ and $\varphi'$. In addition, $t_{nms}$ represents the approximate time response of a driver's steering maneuver through arm muscles (Pick and Cole 2008).

In order to describe the driver interaction and reliance on the haptic guidance steering, the neuromuscular reaction gain for haptic guidance $K_{hg}$ was proposed. The range of $K_{hg}$ (i.e., from 0 to 1) represents the degree of the driver's reliance on the haptic guidance steering. $K_{hg}$ equals 0 indicates that the driver significantly relies on the haptic guidance steering. The weighted visual and haptic guidance by the human driver are represented by $K_d$ and $K_{hg}$. Thereby, the neuromuscular system of the driver model is expressed as

$$K_d \varphi' + K_{nms}(\varphi' - \varphi) - K_{hg} T_h = t_{nms} \dot{T}_d + T_d \qquad (8)$$

### *2.6. Haptic guidance system*

Haptic guidance model is also based on the driver's two-point visual model presented in Fig. 3. The parameters and variables for the haptic guidance system are presented in Table 3. The target trajectory for haptic guidance is set by the centerline of the lane. Because an increase of human driver's derivative action can significantly increase his/her workload, the haptic guidance system is supposed to be effective in compensating for the limitations of human driver derivative action. Considering this, a proportional-derivative control theory is used to manage $e'_y$ of the near point and $e'_\theta$ of the far point. Thereby, it ensures that the vehicle accurately follows the target trajectory. The haptic guidance torque $T_h$ is expressed as

$$T_h = K_1(a'_1 e'_y + a'_2 \dot{e}'_y + a'_3 e'_\theta + a'_4 \dot{e}'_\theta) \qquad (9)$$

Here, $a'_1$, $a'_2$, $a'_3$, and $a'_4$ are constant gains of $e'_y$, $\dot{e}'_y$, $e'_\theta$, and $\dot{e}'_\theta$, respectively; $K_1$ is a constant gain of the overall haptic guidance torque (Wang et al. 2017a).

## 3. Experimental Study

### *3.1. Participants*

Fifteen healthy males participated in the experiment. Experimental data from one participant were omitted owing to poor quality in the measured signal, thereby leaving a sample of fourteen participants for analyzing data and providing results.

Their ages ranged from 21 y to 54 y (mean = 26.6 y, SD = 8.7), and all participants possessed a valid Japanese driver's license for at least one year (mean = 5.4 y, SD = 7.2). At the time of performing driving tasks, all participants had normal or corrected-to-normal vision. Each participant received a monetary reward for participating in the experiment. The experiment content and process were reviewed and approved by the Experiment Ethics Committee of the University of Tokyo (No. 14-113).

### 3.2. Apparatus

As shown in Fig. 4, the experiment was carried out with a high-fidelity driving simulator (Mitsubishi Precision Co., Ltd., Japan). The driving simulator includes a brake pedal and an accelerator pedal, an electric steering system, an instrument panel, and two rearview mirrors. In addition, a driving scene with a 140° field-of-view was visualized by using three projectors.

The driving road was a two-lane expressway with lane width of 3.6 m and the left side was an emergency lane. The lane was marked with solid and broken lines. The participants were instructed to hold the steering wheel at the "Ten" and "Two" positions and to maintain the vehicle in the centerline of the left lane as accurately as possible during the driving task.

In the driving simulator, an electronic steering control system was connected to the host computer via a controller-area network bus. The electronic steering system consisted of a steering wheel, a servomotor, and an electronic control unit (ECU). The haptic guidance torque was calculated by the host computer in real time, and then it was input to the ECU to activate the servomotor that applied the haptic guidance torque to the steering wheel. The limit to the haptic guidance torque was set as 5 N·m (Wang et al. 2017b), so that the driver would be able to override it in all circumstances by exerting more torque on the steering wheel.

The input to the steering system was directly determined by both the driver torque and haptic guidance torque. The target trajectory, which was set by the centerline of the lane, was previously saved in the scenario. The driver's steering torque was measured with a torque sensor at a resolution of 0.005 N·m, and the steering wheel angle was measured with an angular sensor at a resolution of 0.1 degrees. The sampling rate of raw data recorded in the host computer is 120 Hz.

### 3.3. Scenario

As mentioned in the Section "1. Introduction", experimental data are acquired from the driving simulator study (Wang et al. 2019). The driving course is shown in Fig. 5. The driving speed was fixed at 60 km/h through a driving simulator software setting. Thereby, the participant was only asked to operate the steering wheel and operating the gas and brake pedal was not required. This is because the steering performance is highly related to driver's choice of speed, and the model identification of steering behavior would be less difficult by eliminating the speed variability. As a period of time is needed for the participants to adapt their steering behavior to the haptic guidance system, the data recorded between P1 and P2 were used to identify the driver model and the data from first half driving course before P1 were ignored.

## 4. Model Identification and Validation

This section explains how the proposed driver model matches driver behavior with integrated visual and haptic guidance. The measured data in the experimental study were used to develop models for manual driving and driving with haptic guidance.

### 4.1. Parameter identification

The measured data included the vehicle trajectory, haptic guidance torque, steering wheel angle, and driver input torque. The vehicle trajectory and haptic guidance torque were calculated by the host computer connected to the driving simulator. The steering wheel angle was measured by an angular sensor, and the driver input torque was measured by a torque sensor.

By using the lateral error $e_y$, yaw error $e_\theta$, steering wheel angle $\varphi$, and haptic guidance torque $T_h$ as inputs, and driver input torque $T_d$ and target steering wheel angle $\varphi'$ as outputs, the state-space representation of driver model is:

$$\dot{x}(t) = Ax(t) + Bu(t)$$

$$y(t) = Cx(t) + Du(t) + e(t) \qquad (10)$$

$$x(0) = x_0$$

$$\begin{bmatrix} \dot{x}_1 \\ \dot{x}_2 \\ \dot{x}_3 \end{bmatrix} = \begin{bmatrix} 0 & 0 & 0 \\ a_2 \dfrac{2}{t_p} & -\dfrac{2}{t_p} & 0 \\ -a_2 \dfrac{K_d + K_{nms}}{t_{nms}} & \dfrac{2(K_d + K_{nms})}{t_{nms}} & -\dfrac{1}{t_{nms}} \end{bmatrix} \begin{bmatrix} x_1 \\ x_2 \\ x_3 \end{bmatrix}$$

$$+ \begin{bmatrix} 1 & 0 & 0 & 0 \\ a_1 \dfrac{2}{t_p} & a_4 \dfrac{2}{t_p} & 0 & 0 \\ -a_1 \dfrac{K_d + K_{nms}}{t_{nms}} & -a_4 \dfrac{K_d + K_{nms}}{t_{nms}} & -\dfrac{K_{nms}}{t_{nms}} & \dfrac{-K_{hg}}{t_{nms}} \end{bmatrix} \begin{bmatrix} e_y \\ e_\theta \\ \varphi \\ T_h \end{bmatrix}$$

$$\begin{bmatrix} T_d \\ \varphi' \end{bmatrix} = \begin{bmatrix} 0 & 0 & 1 \\ -a_2 & 2 & 0 \end{bmatrix} \begin{bmatrix} x_1 \\ x_2 \\ x_3 \end{bmatrix} + \begin{bmatrix} 0 & 0 & 0 & 0 \\ -a_1 & -a_3 & 0 & 0 \end{bmatrix} \begin{bmatrix} e_y \\ e_\theta \\ \varphi \\ T_h \end{bmatrix} \quad (11)$$

When only the driver torque output is considered, analysis reveals a low model identifiability. Doing so would lead to an identified driver model with a higher degree of fitting, but in the simulation study, the vehicle cannot follow the target trajectory adequately. In that case, $\varphi'$ is taken into account as an additional output in the driver model.

To approximate the time delay $t_p$, a first-order Pade expansion with a rational transfer function is used, given by

$$e^{-t_p s} = \frac{1 - 0.5 t_p s}{1 + 0.5 t_p s} \qquad (12)$$

The discretized state-space realization of the model is:

$$x(k + 1) = Ax(k) + Bu(k)$$

$$y(k) = Cx(k) + Du(k) + e(k) \qquad (13)$$

$$x(0) = x_0$$

The prediction error method implemented in the "Grey Box Identification Toolbox" of MATLAB was used to identify the driver model parameters. The identification iteration stopped when the percentage difference between the loss function's current value and its expected improvement at the next iteration was less than 0.01%. In order to calculate the percentage difference, the Gauss-Newton vector computed for the current parameter value was used to estimate the expected loss function improvement at the next iteration.

The default values and variation intervals for the parameters identification of the driver model, as shown in Table 4, are achieved by a trial-and-error process, and referring to (Saleh et al. 2013, You and Tsiotras 2016).

### 4.2. Identification results

The driver input torque fitting result from a typical participant under the condition of haptic guidance illustrates the comparison of driver input torque between an actual human driver and the identified model; the fitness was 73.7%, as shown in

Fig. 6. As indicated by the identification results in Tables 5 and 6 for the driver model, the proposed driver model matches the driver input torque with an average fitness of 76% for manual driving and an average fitness of 69% under the condition of haptic guidance. The decreased fitness in conditions of haptic guidance as compared to manual driving results from the complicated interactions between the driver and the haptic guidance system. The fitness result demonstrates that this is a satisfying performance, regardless of the inter-subject driver steering behavior.

Noticeable results were found for the 14 participants of the driver model in regards to the identification results of $K_d$ and $K_{hg}$ in the neuromuscular system. The values of $K_d$ under the conditions of manual driving and haptic guidance, as shown in Tables 5 and 6, were compared and we found that the mean value of $K_d$ for haptic guidance was lower than that for manual driving. It indicates that the driver steering effort was reduced by the haptic guidance system, given that $K_d$ represents the steering torque provided by the neuromuscular system in proportion to $\varphi'$, as shown in Fig. 1. This finding was evident for most participants, based on the analysis of $K_d$ for each participant. Furthermore, the comparison of $K_{hg}$ for each participant reveals a tendency for a relatively low value of $K_{hg}$ to correspond to a relatively high degree of driver's reliance on the haptic guidance system, leading to the decrease of $K_d$ for haptic guidance (relative to manual driving). This result is in accordance with the previously-observed reduction of steering effort, if drivers choose to follow the haptic guidance steering system (Abbink et al. 2012).

### *4.3. Validation test*

The model validation test was performed with MATLAB Simulink using the identified parameters of the driver visual and neuromuscular systems, along with the vehicle and steering systems. The longitudinal speed of the vehicle was set to be 60 km/h, corresponding to the driving speed in the experiment. The driving course was identical to the experimental course (P1 to P2), as shown in Fig. 5.

### 4.4. Validation results

Fig. 7 compares the simulated output of the vehicle trajectory against the measured vehicle trajectory, using one typical participant who steered with haptic guidance. It can be observed that the simulated vehicle trajectory followed the target trajectory within the lane along the whole driving course. In addition, the actual driver and the identified model showed similar profiles when negotiating a curve, as shown in Fig. 7b. The absolute mean error along the entire driving course between the measured and simulated vehicle trajectories was only 0.155 m, demonstrating a goodness-of-fit. The absolute mean trajectory error among the 14 participants is between 0.154 m and 0.366 m (M = 0.218 m, SD = 0.061 m), suggesting that the proposed model is capable of predicting driver steering behavior under haptic guidance.

## 5. Model Evaluation

The aim of model evaluation is to use numerical analysis to provide additional evidence to support the capability of the proposed driver model to predict driver behavior.

### 5.1. Declined visual attention

A human operator has a processing time delay before taking action in response to a given stimulus. It has been found that the existence of the time delay $t_p$ is a fundamental cause of unstable vehicle motion and is lengthened by reduced driver inattentiveness (Macadam 2003). Thus, human driving characteristics are simplified here as time delays vary in duration, which are presented in Table 7. Other numerical values of the driver model, haptic guidance system, vehicle dynamics and steering column dynamics are presented in Tables 8 and 9.

Different combinations of $K_d$ and $K_{hg}$ are used to describe different weights to visual and haptic guidance. A lower value of $K_d$ associated with a lower value of $K_{hg}$ indicates a higher degree of driver's reliance on haptic guidance. In the model evaluation studies, high-reliance corresponds to $K_d = 2.0$, $K_{hg} = 0$; mid-reliance corresponds to $K_d = 3.0$, $K_{hg} = 0.5$; low reliance corresponds to $K_d = 4.0$, $K_{hg} = 1.0$. Additionally, manual driving was conducted for comparison and corresponds to $K_d = 4.0$, $T_h = 0$.

The lane keeping accuracy, which reflects driving safety, is measured by the lateral error from the centerline of lane. A lower value of lateral error indicates a higher lane keeping accuracy. The lateral error for different degrees of driver's reliance on haptic guidance was compared among the three drivers, as shown in Fig. 8. The results are from the first curve between P1 and P2 shown in Fig. 5. From the results of Driver 1, it can be observed that there is no evident effect of the haptic guidance system on improving the lane keeping, and moreover, there is no evident difference in the lane keeping accuracy between different degrees of driver's reliance on haptic guidance. In contrast, driving with haptic guidance highly improves the lane keeping accuracy of Driver 3 which has a longer processing time delay (Wang et al. 2017b). A higher degree of driver's reliance leads to a better lane keeping accuracy. The lane keeping accuracy of Driver 2 is somewhere between the performance of Driver 1 and Driver 3. By comparing different degrees of driver's reliance with manual driving, it can be observed that, when the degree of driver's reliance on haptic guidance is low, the driver's lane keeping performance is closer to that in manual driving (Abbink et al. 2012).

## *5.2. Haptic guidance system failure*

In order to further evaluate the effect of different degrees of driver's reliance on the haptic guidance system, a system failure was addressed in the numerical simulation. When a system failure occurs, the haptic guidance system will stop providing active torques on the steering wheel.

In detail, the failure of the haptic guidance system occurs a certain time (70 seconds) after the starting point of the driving, thereby indicating that the failure occurs at the first road curve between P1 and P2 of the driving course as shown in Fig. 5. After the failure of system occurs, the remained part would be manually driving as active assistance torque is not provided. Moreover, it is assumed that the driver needs a certain time (one second in this study) to respond to the failure and to change the driving characteristics. It means that the parameter values of the driver model remain the same as in the haptic guidance condition within the one second after the failure occurrence. After one second of driver's response time, the parameter values of the driver model become identical to those used in the condition of manually driving.

The lateral error was measured to evaluate the lane keeping accuracy when a system failure occurs. The lateral error for different degrees of driver's reliance on haptic guidance was compared among the three drivers, as shown in Fig. 9. The failure of haptic guidance system leads to a sudden increase of lateral error on the three drivers, and even larger lateral error is yielded by the higher degree of driver's reliance on the system (de Winter and Dodou 2011). This suggests that a downside to the increase of the lane keeping accuracy by haptic guidance is a sudden decrease of driving performance when a failure of system occurs. The comparison of differences in attentiveness indicates that the driver with lower visual attention might prefer a higher degree of reliance on haptic guidance. As for a cost-benefit analysis considering the system failure, it is assumed that a driver might balance the increase of the lane keeping accuracy with an avoidance of a sudden performance decrement by adjusting the degree of reliance on haptic guidance.

## 6. Discussion

From the model identification results regarding the goodness-of-fit to the measured driver torque input, we found that the proposed driver model matches the driver input torque with an average fitness of 76% for manual driving and an average fitness of 69% under the condition of haptic guidance. These fitness values are acceptable, given the individual variability in steering behavior. The results of the lane-following performances in the first part of the experimental study indicate that the individual variability in steering behavior was quite large. Individual differences such as driving experience, age, or physical factors like body size could explain the individual variability in steering behavior. Although the participants were instructed to driver as close to the centerline of lane as possible, drivers always naturally cut curves in the reality, which could also cause a relatively lower fitness. To improve the model fitness, future studies could employ a more advanced nonlinear driver parameter estimation method. For example, a nonlinear Kalman filter has been previously used to understand the steering behavior of different types of drivers (You et al. 2017).

The model validation and evaluation results suggest that a driver model with a small number of parameters, especially with the weighting process for $K_d$ and $K_{hg}$, is capable of predicting driver interactions with the haptic guidance system. A driver model with a small number of parameters is beneficial in energy saving and response time when building an online driver model for real-vehicle development processes. There has been work in human-robot interaction, e.g., a human adaptation model was built online for designing a mutually-adaptive human-robot cooperation system (Nikolaidis et al. 2017). For driver-automation interaction, it is expected that a dynamic weighting process for visual and haptic guidance will lead to a better understanding of driver adaptation behavior, and further lead to an optimum design for haptic guidance systems. Moreover, the neuromuscular reflex gain $K_{nms}$ is fixed in the current model, whereas some researchers proposed that $K_{nms}$ could reflect the muscular co-contraction, which is a way for the driver to increase the stiffness of system (Pick and Cole 2008, Mars et al. 2011). In the future study, the effect of variation of $K_{nms}$ will be addressed.

The current driver model consists of a visual system and a neuromuscular system, which are linear and time-invariant. In the current experiment, the driver steering behavior could be considered as approximately constant throughout the driving trial in the lane-following task, as the driving simulator experiment was carefully designed. A driver's cognition, which is crucial for higher-level driving behavior such as decision making and planning (Salvucci et al. 2001), could improve the driver model. However, it was not considered in the current model. In future studies, advanced non-linear modeling that considers the driver's cognition, e.g., fuzzy modeling, could be explored to address higher-level driver behavior.

Another issue concerning the model is the driver feedback from visual guidance during a lane-following task. There is still controversy over how visual information is used by drivers (Wallis et al. 2007). In a numerical simulation case study with a condition of declined visual attention, there is a limitation in simply representing the declined visual attention with a higher value of the processing time delay. It has been found that the processing time delay increases when the driver visual attention declines owing to fatigue or distracted driving. However, actual driver behavior will be more complicated. According to (You et al. 2017), a processing time delay for a normal driver would be between 0.01 to 0.3 s. A higher value of 0.5 s was used in the simulation study to represent a driver with declined visual attention. Considering inter-subject differences in driver behavior, additional parameters will be addressed in future studies.

In addition to the driver's cognitive state, driver behavior will also be influenced by environmental factors in real driving situations. For example, other vehicles arrive at various intervals, and the driver would have to consider the intentions of other drivers. Traffic signs on the road could also attract the driver's attention. Thus, the current driver model is a simplified one that basically deals with a lane-following task in a monotonous driving environment. Thus, the driver model should address additional environmental factors for a real-life driving task in future work.

A further limitation of the present study was that the sample was biased toward male drivers, and female drivers should be addressed in future work. Owing to the above limitations, it would be premature to draw broad conclusions, as further validation of the proposed model is still required. However, within the scope of this study, it has been shown that the proposed driver model is capable of predicting driver behavior, with the potential to become a useful tool for optimally designing and evaluating haptic guidance.

## 7. Conclusion

This paper proposed a novel driver model through a weighting of visual guidance and haptic guidance for a lane-following task when driving with a steering assistance system. The weighting process was developed to address the interaction and reliance of the driver in relation to the haptic guidance steering.

On the basis of collected data from driving simulator experiment, the proposed driver model was proved to be suitable for the identification of steering behavior with haptic guidance. In particular, the driver input torque was matched by the driver model with an average fitness of 69% across the fourteen participants. Additionally, the identification results indicate that a higher degree of driver reliance on the haptic guidance system would lead to in a decrease in driver's steering effort. Moreover, the validation results show that the simulated trajectory effectively followed the driving course by matching the measured trajectory from the experiment. The model evaluation by numerical simulation suggests that the parameterized driver model, especially with respect to the model gains $K_d$ and $K_{hg}$, is capable of predicting driver behavior in the case of inter-subject variability in driver attentiveness as well as in the case of a system failure.

Finally, the observations in this paper reveal the potential of the proposed model to be used for conducting numerical simulations and on-line model building for further design and evaluation of a haptic guidance system. As for future work, female participants will be added in the follow-up study.

Table 1. Parameters and Variables for Vehicle Dynamics and Steering System

|  | Definition |
|---|---|
| $B_s$ | Damping of steering system |
| $E_t$ | Sum of pneumatic and castor trail |
| $I$ | Yaw moment inertia of vehicle |
| $J_s$ | Moment inertia of steering system |
| $K_f$ | Cornering stiffness of front tire |
| $K_r$ | Cornering stiffness of rear tire |
| $K_s$ | Spring constant converted around the kingpin |
| $K_t$ | Transmission ratio between steering wheel angle and front wheel steering angle |
| $l_f$ | Longitudinal position of front wheels to center of gravity of vehicle |
| $l_r$ | Longitudinal position of rear wheels to center of gravity of vehicle |
| $m$ | Mass of vehicle |
| $r$ | Yaw rate |
| $T_a$ | Aligning torque |
| $v$ | Vehicle speed |
| $\beta$ | Side slip angle |
| $\delta$ | Front wheel steering angle |
| $\varphi$ | Steering wheel angle |

Table 2. Parameters and Variables for Driver Model

| | Definition |
|---|---|
| $a_1$ | Constant gain for $e_y$ |
| $a_2$ | Constant gain for integral of $e_y$ |
| $a_3$ | Constant gain for $e_\theta$ |
| $e_y$ | Lateral error of near point for driver model |
| $e_\theta$ | Yaw error of far point for driver model |
| $K_d$ | Target steering angle to torque gain |
| $K_{hg}$ | Neuromuscular reaction gain for haptic guidance |
| $K_{nms}$ | Neuromuscular reflex gain |
| $t_f$ | Look-ahead time of far point for driver model |
| $t_n$ | Look-ahead time of near point for driver model |
| $t_{nms}$ | Neuromuscular time constant |
| $t_p$ | Processing time delay by human driver |
| $T_d$ | Driver input torque |
| $\varphi'$ | Target steering wheel angle |

Table 3. Parameters and Variables for Haptic Guidance System

| | Definition |
|---|---|
| $a'_1$ | Constant gain for $e'_y$ |
| $a'_2$ | Constant gain for derivative of $e'_y$ |
| $a'_3$ | Constant gain for $e'_\theta$ |
| $a'_4$ | Constant gain for derivative of $e'_\theta$ |
| $e'_y$ | Lateral error of near point for haptic guidance model |
| $e'_\theta$ | Yaw error of far point for haptic guidance model |
| $K_l$ | Constant gain for haptic guidance torque |
| $t'_f$ | Look-ahead time of far point for haptic guidance model |
| $t'_n$ | Look-ahead time of near point for haptic guidance model |
| $T_h$ | Haptic guidance torque |

Table 4. Driver Model Parameters

|          | Default value | Variation interval |
|----------|---------------|--------------------|
| $a_1$    | 0.1           | [0-0.5]            |
| $a_2$    | 0.01          | [0-0.1]            |
| $a_3$    | 3.7           | [3-5]              |
| $t_p$    | 0.1           | [0.01-0.3]         |
| $K_d$    | 3             | [1-5]              |
| $K_{lg}$ | 0.5           | [0-1]              |
| $K_{nms}$| 1             |                    |
| $t_{nms}$| 0.1           |                    |

Table 5. Driver Model Identification in the Condition of Manual Driving

| Sub | $a_1$ | $a_2$ | $a_3$ | $t_p$ | $K_d$ | $K_{hg}$ | Fitness |
|-----|-------|-------|-------|-------|-------|----------|---------|
| 1 | 0.068 | 0.029 | 3.699 | 0.090 | 3.774 | - | 78.000 |
| 2 | 0.060 | 0.015 | 3.442 | 0.169 | 3.506 | - | 72.395 |
| 3 | 0.066 | 0.019 | 3.656 | 0.014 | 3.861 | - | 79.452 |
| 4 | 0.097 | 0.019 | 3.759 | 0.041 | 3.809 | - | 76.801 |
| 5 | 0.082 | 0.011 | 3.374 | 0.297 | 3.646 | - | 72.383 |
| 6 | 0.081 | 0.009 | 3.502 | 0.300 | 3.674 | - | 73.455 |
| 7 | 0.053 | 0.023 | 3.530 | 0.027 | 3.759 | - | 74.188 |
| 8 | 0.075 | 0.026 | 3.729 | 0.023 | 3.947 | - | 81.110 |
| 9 | 0.080 | 0.016 | 3.744 | 0.120 | 3.649 | - | 73.171 |
| 10 | 0.085 | 0.014 | 3.428 | 0.090 | 3.732 | - | 77.894 |
| 11 | 0.049 | 0.029 | 3.561 | 0.057 | 3.856 | - | 79.559 |
| 12 | 0.078 | 0.029 | 3.298 | 0.086 | 3.794 | - | 77.215 |
| 13 | 0.074 | 0.015 | 3.523 | 0.038 | 3.719 | - | 77.653 |
| 14 | 0.066 | 0.020 | 3.627 | 0.010 | 3.981 | - | 79.143 |

Table 6. Driver Model Identification in the Condition of Haptic Guidance

| Sub | $a_1$ | $a_2$ | $a_3$ | $t_p$ | $K_d$ | $K_{hg}$ | Fitness |
|-----|-------|-------|-------|-------|-------|----------|---------|
| 1 | 0.100 | 0.029 | 3.781 | 0.034 | 3.249 | 0.526 | 69.099 |
| 2 | 0.120 | 0.027 | 3.650 | 0.300 | 2.888 | 0.393 | 68.291 |
| 3 | 0.047 | 0.025 | 3.601 | 0.300 | 2.781 | 0.421 | 52.413 |
| 4 | 0.085 | 0.025 | 3.803 | 0.184 | 3.260 | 0.509 | 72.441 |
| 5 | 0.060 | 0.023 | 3.637 | 0.229 | 4.056 | 0.926 | 73.700 |
| 6 | 0.111 | 0.048 | 3.711 | 0.010 | 2.300 | 0.002 | 66.319 |
| 7 | 0.053 | 0.014 | 3.672 | 0.300 | 3.859 | 0.783 | 72.804 |
| 8 | 0.049 | 0.025 | 3.692 | 0.017 | 3.992 | 0.971 | 73.317 |
| 9 | 0.058 | 0.014 | 3.694 | 0.300 | 2.492 | 0.137 | 63.885 |
| 10 | 0.067 | 0.023 | 3.517 | 0.124 | 3.227 | 0.392 | 73.111 |
| 11 | 0.061 | 0.029 | 3.566 | 0.300 | 3.593 | 0.615 | 69.442 |
| 12 | 0.066 | 0.033 | 3.490 | 0.010 | 2.387 | 0.008 | 71.005 |
| 13 | 0.086 | 0.003 | 3.476 | 0.046 | 2.125 | 0.035 | 66.517 |
| 14 | 0.047 | 0.011 | 3.620 | 0.018 | 3.945 | 0.858 | 73.599 |

Table 7. Three Driver Model Parameters

| | Processing time delay (s) |
|---|---|
| *Driver 1* | 0.1 |
| *Driver 2* | 0.3 |
| *Driver 3* | 0.5 |

Table 8. Values of Driver Model and Haptic Guidance Systems for Numerical Simulation

| | Value | Unit |
|---|---|---|
| $a_1$ | 0.1 | - |
| $a_2$ | 0.05 | - |
| $a_3$ | 3.7 | - |
| $t_f$ | 1.0 | s |
| $t_n$ | 0.3 | s |
| $t_{nms}$ | 0.1 | s |
| $K_{nms}$ | 1.0 | - |
| $a'_1$ | 2 | - |
| $a'_2$ | 0.05 | - |
| $a'_3$ | 40 | - |
| $a'_4$ | 1 | - |
| $K_1$ | 0.25 | - |
| $t'_f$ | 0.7 | s |
| $t'_n$ | 0.3 | s |

Table 9. Values of Vehicle Dynamics and Steering Column Dynamics for Numerical Simulation

|  | Value | Unit |
|---|---|---|
| $v$ | 60 | km/h |
| $m$ | 1100 | kg |
| $I$ | 2940 | kg·m$^2$ |
| $l_f$ | 1 | m |
| $l_r$ | 1.635 | m |
| $K_f$ | 53300 | N/rad |
| $K_r$ | 117000 | N/rad |
| $K_s$ | 48510 | N·m/rad |
| $E_t$ | 0.026 | - |
| $B_s$ | 0.57 | N·m·s/rad |
| $J_s$ | 0.11 | kg·m$^2$ |
| $K_t$ | 1/17 | - |

Figure 1. Driver-vehicle-road model with haptic guidance system.

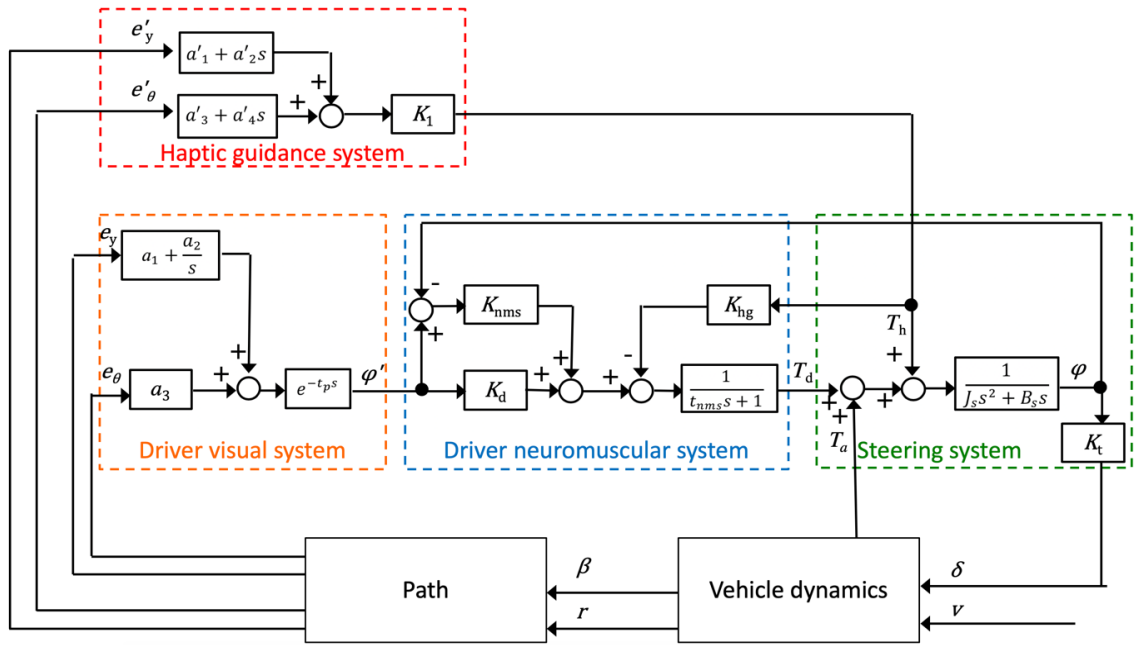

Figure 2. Equivalent bicycle model.

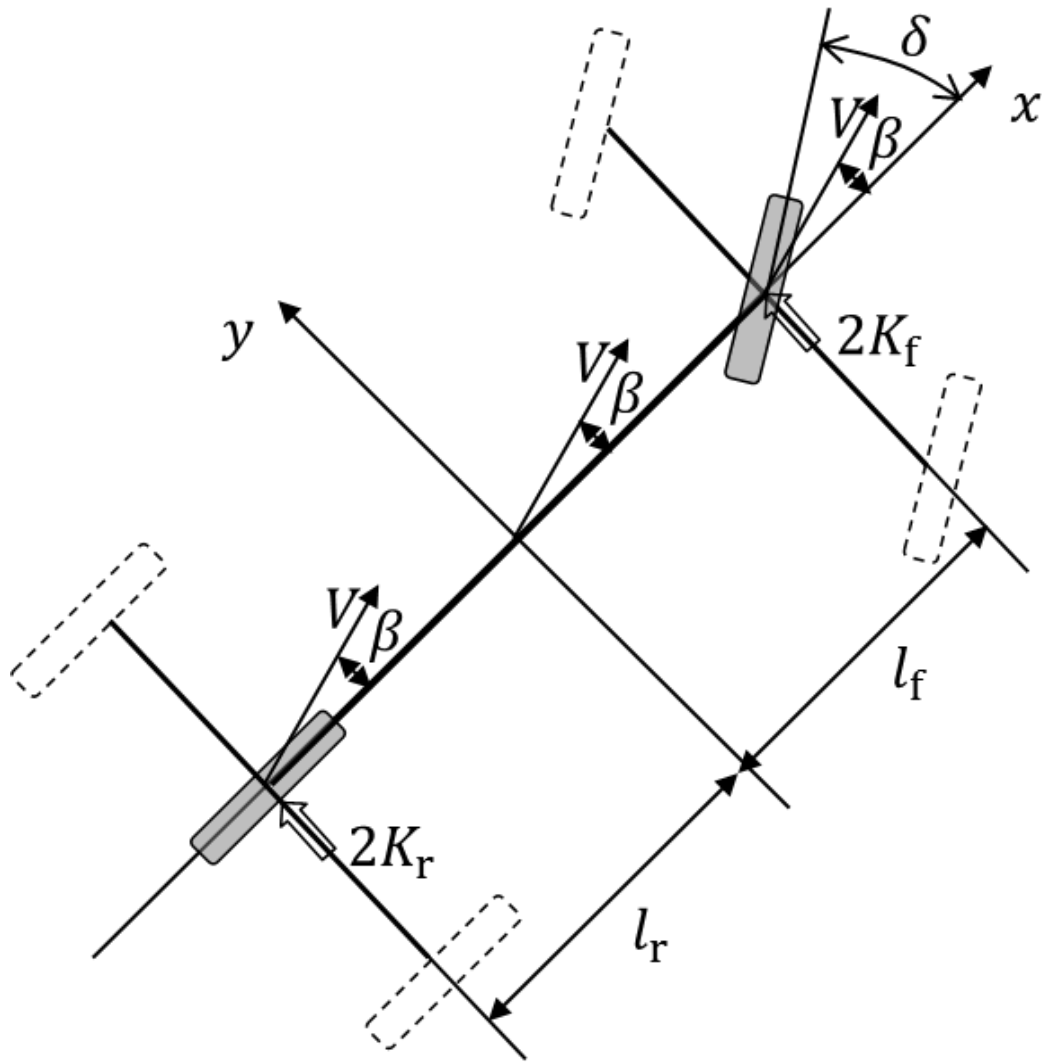

Figure 3. The geometric relationship between the road and driver's two-point visual model.

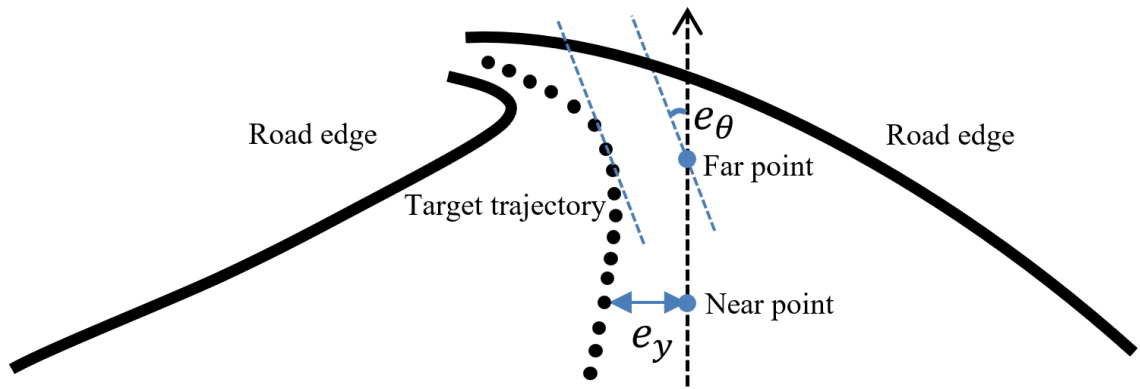

Figure 4. Driving simulator and environment in the experiment.

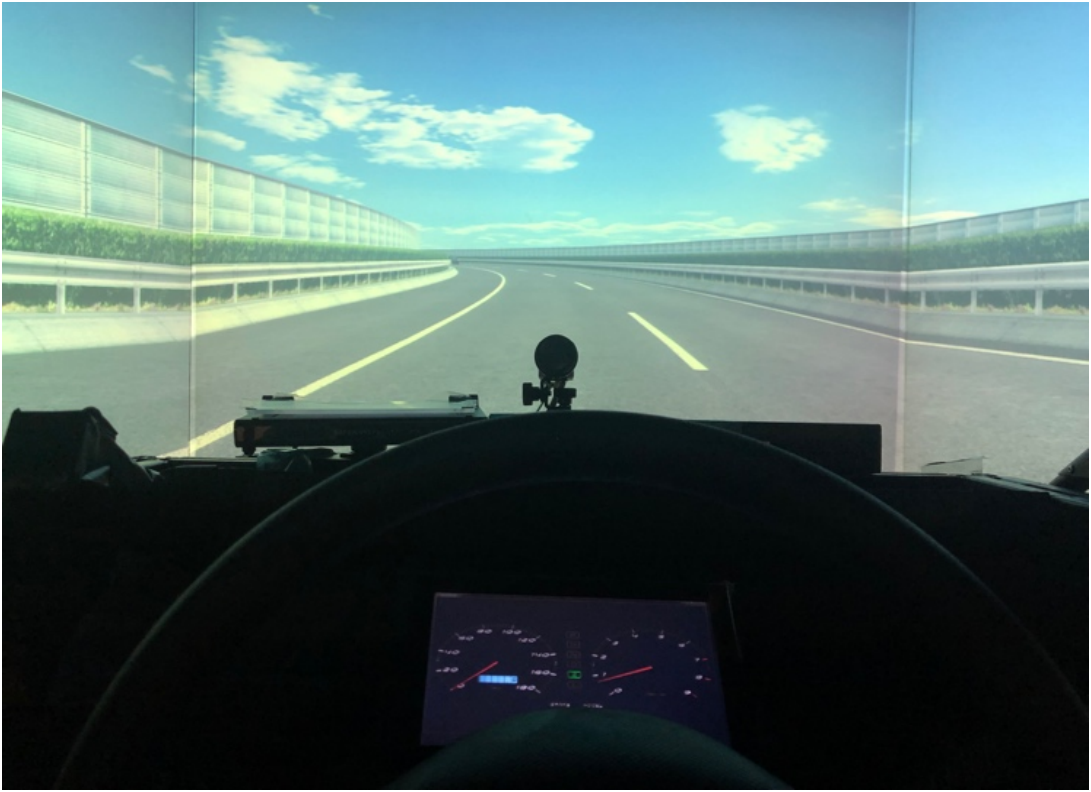

Figure 5. Driving course in the driving simulator experiment.

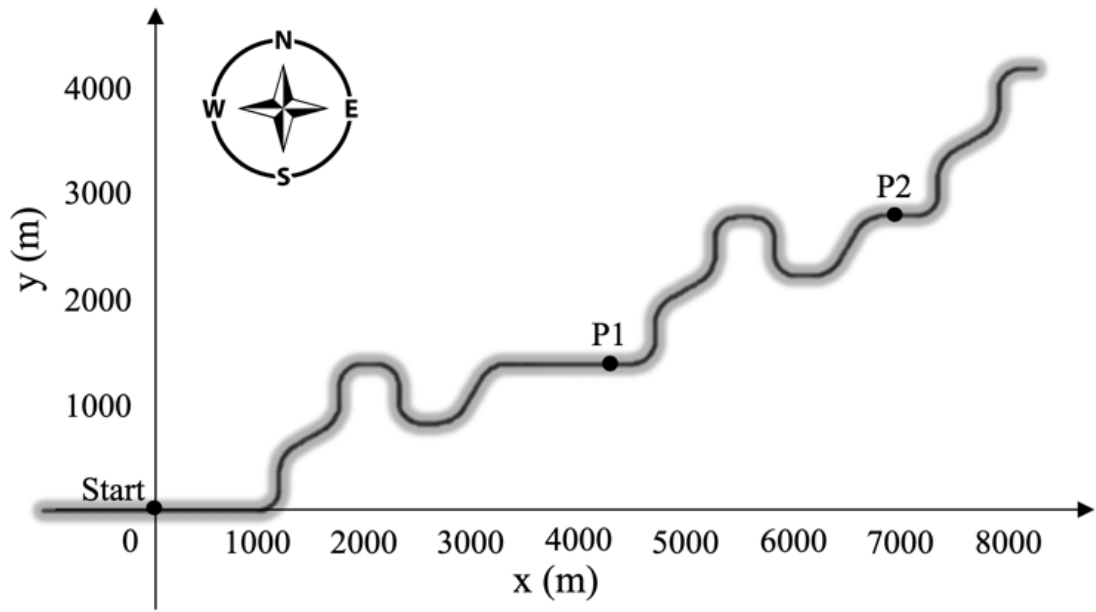

Figure 6. An example of driver steering input torque fitting in the condition of haptic guidance.

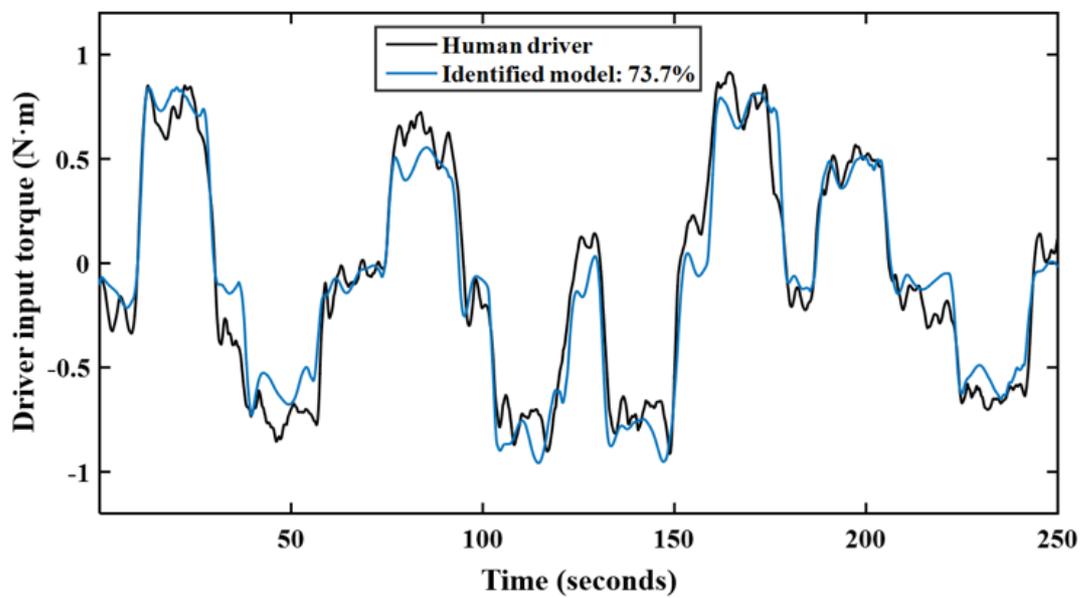

Figure 7. An example of comparison in vehicle trajectory between measured and simulated results along the driving course in the condition of haptic guidance; (a) entire driving course (b) a part of the driving course.

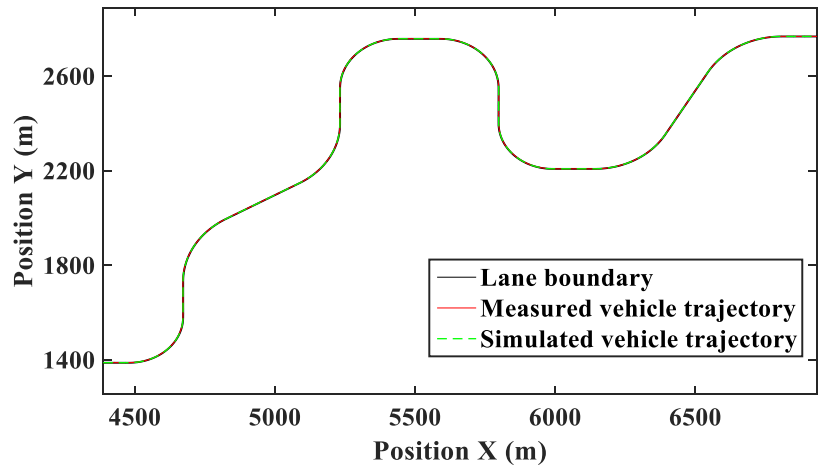

(a)

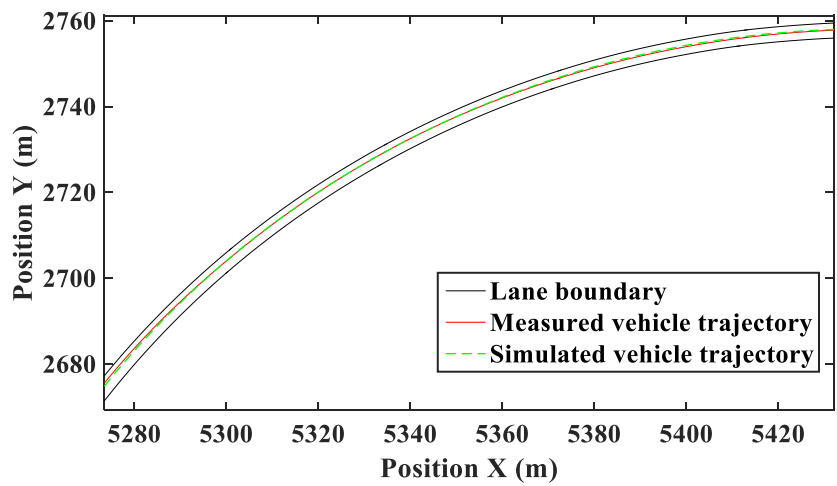

(b)

Figure 8. Lateral error from centerline of lane for different degrees of driver's reliance on haptic guidance; (a) Driver 1, (b) Driver 2, and (c) Driver 3.

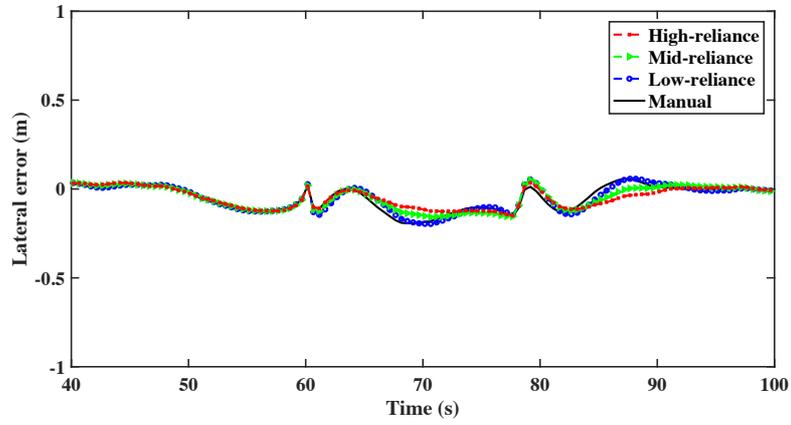

(a)

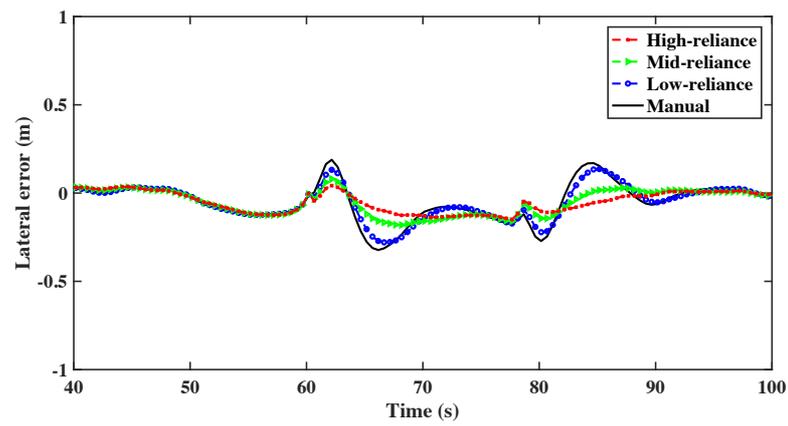

(b)

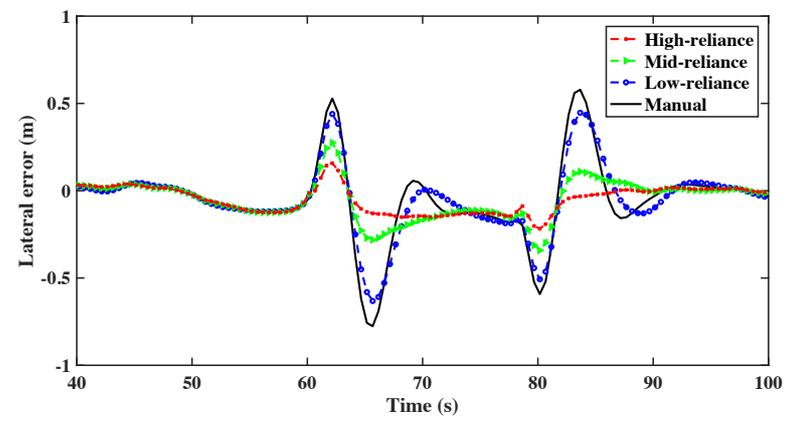

(c)

Figure 9. Lateral error from centerline of lane with system failure; (a) Driver 1, (b) Driver 2, and (c) Driver 3.

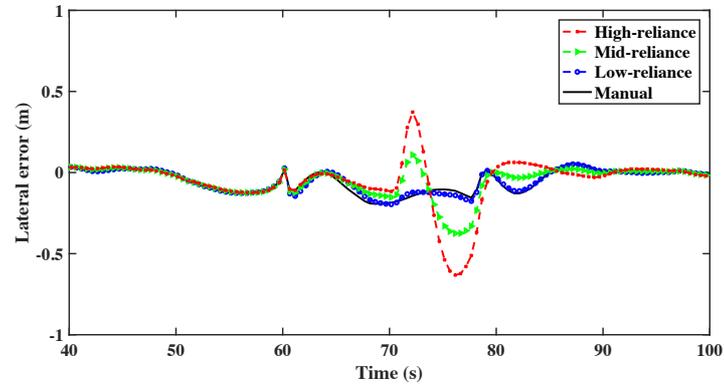

(a)

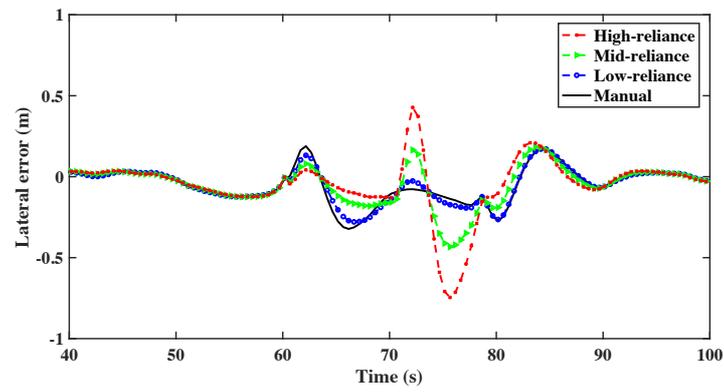

(b)

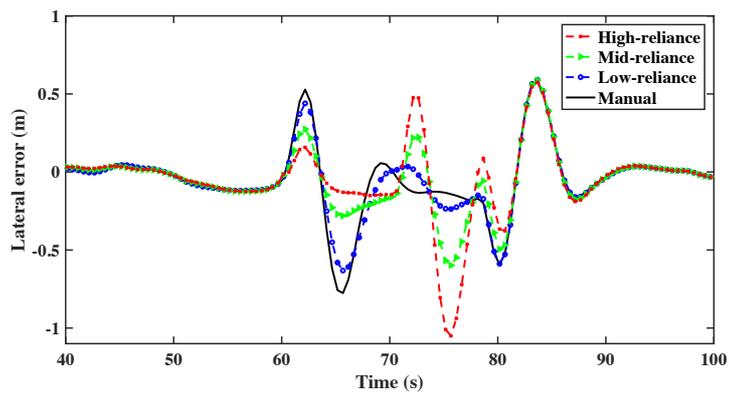

(c)

# A list of figure captions

Figure 1. Driver-vehicle-road model with haptic guidance system.

Figure 2. Equivalent bicycle model.

Figure 3. The geometric relationship between the road and driver's two-point visual model.

Figure 4. Driving simulator and environment in the experiment.

Figure 5. Driving course in the driving simulator experiment.

Figure 6. An example of driver steering input torque fitting in the condition of haptic guidance.

Figure 7. An example of comparison in vehicle trajectory between measured and simulated results along the driving course in the condition of haptic guidance; (a) entire driving course (b) a part of the driving course.

Figure 8. Lateral error from centerline of lane for different degrees of driver's reliance on haptic guidance; (a) Driver 1, (b) Driver 2, and (c) Driver 3.

Figure 9. Lateral error from centerline of lane with system failure; (a) Driver 1, (b) Driver 2, and (c) Driver 3.